\documentstyle[12pt,aaspp4]{article}

\begin{document}

\def\kms{km~s$^{-1}$}
\def\msun{$M_{\odot}$}
\def\rsun{$R_{\odot}$}
\def\lsun{$L_{\odot}$}
\def\halpha{H$\alpha$}
\def\hbeta{H$\beta$}
\def\hgama{H$\gamma$}
\def\hdelta{H$\delta$}
\def\Teff{T$_{eff}$}
\def\logg{$log_g$} 

\tighten
%\singlespace
%\doublespace

%\received{}
%\revised{}
%\accepted{}
%\journalid{}{}
%\articleid{}{}
%\slugcomment{\bf DRAFT OF 1/19/99}
\slugcomment { Ap.J. Letters, in press } 
%\slugcomment {Proofs and offprint requests to: James Liebert}

\title{ON THE NATURE OF THE PECULIAR HOT STAR 
IN THE YOUNG LMC CLUSTER NGC~1818 }

\author{James Liebert }
\affil{Steward Observatory}  
\affil{University of Arizona, Tucson AZ 85721}
\affil{liebert@as.arizona.edu}  

\author{and}

\affil{Space Telescope Science Institute}
\affil{3700 San Martin Dr.} 
\affil{Baltimore, MD 21218}

\begin{abstract}

The blue star reported in the field of the young LMC cluster NGC~1818
by Elson et al. (1998) has the wrong luminosity and radius to be a
``luminous white dwarf'' member of the cluster.  In addition, unless
the effective temperature range quoted by the authors is a drastic
underestimate, the luminosity is much too low for it to be a cluster
member in the post-AGB phase.  Other possibilities, including that of
binary evolution, are briefly discussed.  However, the implication
that the massive main sequence turnoff stars in this cluster can
produce white dwarfs (instead of neutron stars) from single-star
evolution needs to be reconsidered.

\end{abstract}

\vfill
\eject

\section{INTRODUCTION}

Elson et al. (1998) -- see also the STScI Press Release of 9 April
1998; http//oposite.stsci.edu/pubinfo/pr/1998/16 -- have announced the
discovery of ``a luminous white dwarf'' in the young star cluster
NGC~1818 in the Large Magellanic Cloud.  The cluster has a main
sequence turnoff mass of between 7.6 and 9.0\msun, depending on
whether convective core overshooting is assumed in the models.  That
an even more massive progenitor might have produced a white dwarf
would be of potential importance in the determination of the upper
limit in initial mass for producing this kind of stellar remnant, and
perhaps the lower limits of neutron stars and/or supernovae.  Of the
previously studied young clusters in the Milky Way Galaxy by Koester
\& Reimers (1996, and references therein), the highest derived initial
stellar mass forming a massive white dwarf is 6.97\msun, and the error
bars for this determination are unfortunately large.

Elson et al. found a candidate bluer than the NGC~1818 main sequence
in their color magnitude diagrams, whose colors suggested an extremely
young (hot) white dwarf. Admittedly it lay outside the cluster core
radius, but they pointed out that red giants of the cluster are also
found outside the cluster core.  Moreover, they argued that the
probability of finding a quasar in this small field was 10$^{-3}$.
Finally, in a "Note added in proof" a spectrum was mentioned that
confirms that the object is a star, and suggests a T$_{eff}$ of
25,000--35,000~K.

\section{A DIFFERENT INTERPRETATION}

\subsection{Not a White Dwarf Cluster Member} 

The designation of this object as a ``white dwarf,'' even in the title
of the paper, is the first issue to be questioned.  A white dwarf is
defined as an object whose interior is characterized by an electron
degenerate equation of state, out to nearly the surface.  They are
known to have radii of the order 10$^{-2}$ solar.  The most luminous
white dwarfs are very hot and of relatively low mass (large radius).
Of over 300 hot white dwarfs analyzed from the Palomar Green Survey
(Liebert et al. 1995), the most luminous in visual magnitude units are
still fainter than M$_V\sim$6 (but are also much hotter than
35,000~K).  At the distance of the Large Magellanic Cloud this would
correspond to an apparent V magnitude of 24.5, compared to V$\sim$18.4
for the NGC~1818 candidate.  At the distance of the Large Magellanic
Cloud (m-M $\sim$18.5), the candidate has an absolute visual magnitude
(M$_V$) near zero.  Such an object with a 30,000~K effective
temperature has an implied radius of the order of solar.

The real inconsistency in radius becomes worse, if the object is
supposed to be a white dwarf formed from the evolution of a massive
(intermediate mass) star.  It is well known (Weidemann 1990) that
clusters with main sequence turnoff masses $\ga$5.0\msun\ produce
massive white dwarfs $\ga$0.9\msun. Thus, due to their abnormally
small radii, the young white dwarfs found in NGC~2516 by Koester \&
Reimers (1996) and other young clusters have absolute visual
magnitudes (M$_V$) no brighter than 10.75.  At the LMC distance, this
would correspond to an apparent V magnitude of $\ga$29.  Note also --
since this is relevant to the following subsection -- that the
estimated masses of the several white dwarfs analyzed in this cluster
span 0.85-1.31\msun.  Perhaps the best analyzed case (because it is
nearest and brightest) is LB~1497 (0349+247), a member of the Pleiades
cluster which also has a turnoff mass of $\ga$5.0\msun; its mass is
found to be 1.025\msun\ based on the gravitational redshift (Wegner,
Reid \& McMahon 1989), or 1.084\msun\ from a model atmospheres
analysis by Bergeron, Liebert \& Fulbright (1995), similar to those of
Koester \& Reimers (1996) .

\subsection{Not a Post-AGB Cluster Member}

One must therefore assume that what Elson et al. (1998) really meant
was that their NGC~2818 candidate is an object on its way to {\it
becoming} a white dwarf -- a post-asymptotic giant branch (post-AGB)
star.  This hypothesis would leave intact their basic conclusion that
this cluster of high turnoff mass stars can produce white dwarfs.  We
note from the previous subsection that, if NGC~2818 were to produce a
post-AGB star, its mass should be high ($\ga$0.9\msun) compared to
typical (older) stellar remnants.  Unfortunately, unless the
25,000--35,000~K temperature range from a spectrum in the ``note
added'' comment is a drastic underestimate, I must argue that the star
cannot be a post-AGB member of the cluster.

It has been well known since the work of Paczynski (1971) that the
luminosity of a post-AGB star increases rapidly with the core mass --
see Iben \& Renzini (1983) for a still-timely review.  A number of
similar calculations (eg. Sch\"onberner 1979, Wood \& Faulkner 1986)
have verified this correlation.  Bl\"ocker \& Sch\"onberner (1991) and
Bl\"ocker (1995) showed that the post-AGB luminosity for a given mass
has some dependence on the structure of the AGB model.  The recent
calculations nonetheless appear to offer similar predictions for the
luminosity of post-AGB cores near 0.9\msun.  For example, the
0.836\msun\ track for solar composition shown in Bl\"ocker (1995) has
log L/\lsun~$\sim$4.25 at 30,000~K as it evolves at nearly constant
luminosity to very high effective temperatures.  A 0.855\msun\ track
for Z=0.004 by Vassiliadis \& Wood (1994) has log L/\lsun~$\sim$4.30.
Yet the Elson et al. cluster candidate has M$_V$ near zero.  At
$T_{eff}$ near 30,000~K the bolometric correction (BC$_V$) is
approximately -3 (Wesemael et al. 1980, for a log~g = 4, pure hydrogen
atmosphere).  M$_{bol}$ varies slowly with $T_{eff}$ and is therefore
not terribly sensitive to the uncertain temperature.  The estimated
luminosity of the candidate at the cluster distance is therefore only
log L/\lsun~$\sim$3.0.  This value appears to correspond to a core
mass of roughly 0.5\msun, requiring extrapolation of published core
mass -- luminosity relations, and arguably too low for a star to even
reach the AGB.

If, on the other hand, the true effective temperature of the NGC~1818
candidate has been drastically underestimated (the ``note added''
remark) and $T_{eff}$ approached 100,000~K, the BC$_V$ might become
large enough for the luminosity to match a massive post-AGB track.
For this to be the case, however, any hydrogen lines detected in the
spectrum would be extremely weak, since most hydrogen would be
ionized.

\subsection{Possible alternative solutions} 

One might conclude that the more likely hypothesis is that the object
is a foreground hot star of the Galactic halo.  An extended horizontal
branch (or hot subdwarf) star is a possible explanation: such objects
may have T$_{eff}$ near 30,000~K, are in the long-lived phase of core
helium-burning, and are characteristic of the metal-poor, halo
population.  If this interpretation is correct, M$_V$ could be more
like +4--5, and the the object might be 5--10 kpc distant.  

The authors state, however, that their measured radial velocity is
consistent with LMC membership, and renders unlikely the possibility
that their candidate could be a foreground halo star.  I conclude with
a few remarks about the implications of this possibility. In
particular, this would mean that a young cluster can produce an
unexpected kind of evolved object that is underluminous compared with
the post-AGB tracks expected for massive stars.

I speculate that binary evolution might provide a solution.  Low mass
white dwarfs of $\la$0.5\msun\ with interiors apparently composed of
helium have been found as companions to stars ranging from low mass
main sequence stars to white dwarfs (Marsh, Dhillon \& Duck 1995) and
millisecond pulsars (Lundgren et al. 1996).  What appears to be
required is that the progenitor of the white dwarf, during post-main
sequence evolution, transfers its envelope to the companion (or loses
it) before the mass of the core reaches the amount required for
ignition of helium.  When no envelope remains, the undermassive
progenitor core could leave the red giant branch (RGB), and evolve on
a track that is parallel to, but at much lower luminosity than, the
post-AGB tracks discussed earlier.  The post-RGB tracks evolve much
more slowly than the massive stellar tracks (D'Cruz et al. 1996),
offering a higher probability of catching a star in this
otherwise-rare phase of evolving to a white dwarf.

Were the speculative hypothesis posed in the previous paragraph to be
proven correct, it would mean that Elson et al.'s (1998) candidate
{\it is} becoming a white dwarf in this cluster with a turnoff mass
between 7.6 and 9.0\msun.  It would {\it not}, however, support the
stated implications for the upper mass limit of white dwarf formation
(and lower limit for neutron star production) from single-star
evolution.  If a low-mass remnant core can form from interacting
binary star evolution, such objects are likely to be rare.

\acknowledgments { } This work was supported by the National Science
Foundation through grant AST92-17961.


\begin{references}

\overfullrule=0pt

\reference{} Bl\"ocker, T. 1995, \aap, 299, 755 

\reference{} Bl\"ocker, T., \& Sch\"onberner, D. 1991, \aap, 240, L11 

\reference{} D'Cruz, N.L., Dorman, B., Rood, R.T., \& O'Connell, R.W. 
1996, \apj, 466, 359

\reference{} Elson, R.A., Sigurdsson, S., Hurley, J., Davies, M.B., \& 
Gilmore, G.F. 1998, \apjl, 499, L53 

\reference{} Iben, Jr., I., \& Renzini, A. 1983, \araa, 21, 271 

\reference{} Koester, D., \& Reimers, D. 1996, \aap, 313, 810 

\reference{} Liebert, J., \& Bergeron, P. 1995, in {\it White 
Dwarfs}, ed. D. Koester \& K. Werner (Heidelberg: Springer), 
p. 12 

\reference{} Lundgren, S.C., Cordes, J.M., Foster, R.S., 
Wolszczan, A. \& Camilo, F. 1996, \apjl, 458, L33

\reference{} Marsh, T.R., Dhillon, V.S., \& Duck, S.R. 1995, 
\mnras, 275, 828 

\reference{} Paczynski, B. 1971, {\it Acta. Astr.}, 21, 417 

\reference{} Vassiliadis, E., \& Wood, P.R. 1994, \apjs, 92, 125 

\reference{} Weidemann, V. 1990, \araa, 28, 103 

\reference{} Wesemael, F., Auer, L.H., Van Horn, H.M., \& 
Savedoff, M.P. 1980, \apjs, 43, 159

\reference{} Wood, P.R., \& Faulkner, D.J. 1986, \apj, 307, 659

\end{references}
\end{document}